\documentclass[print]{revtex4}
\usepackage{amsmath,amssymb}
\usepackage{graphicx}
\newcommand{\lvec}[1]{|#1\!\!>}

\draft

\begin{document}

\title{Technique of quantum state transfer for a double $Lambda$ atomic beam}

\author{Xiong-Jun Liu$^{a,b}$\footnote{Electronic address: x.j.liu@eyou.com}, Hui Jing$^{c}$,
Xiao-Ting Zhou$^d$ and Mo-Lin Ge$^{a,b}$} \affiliation{a.
Theoretical Physics Division, Nankai Institute of
Mathematics,Nankai University, Tianjin 300071, P.R.China\\
 b. Liuhui Center for Applied Mathematics, Nankai
University and Tianjin University, Tianjin 300071, P.R.China\\
 c. Shanghai Institure of Optics and Fine Machines,
CAS, Shanghai 201800, P.R.China\\
 d. College of Physics, Nankai University, Tianjin
300071, P.R.China}

\begin{abstract}
The transfer technique of quantum states from light to collective
atomic excitations in a double $\Lambda$ type system is extended
to matter waves in this paper, as a novel scheme towards making a
continuous atom laser. The intensity of the output matter waves is
found to be determined by the initial relative phase of the two
independent coherent probe lights, which may indicate an
interesting method for the measurement of initial relative phase
of two independent light sources.\\

\end{abstract}
\maketitle

\indent Recently, the novel mechanism of Electromagnetically
Induced Transparency (EIT) \cite{1} and its many important
applications have attracted much attention in both experimental
and theoretical aspects \cite{2,3,4}, especially after
Fleischhauer and Lukin \cite{5} proposed their famous dark-states
polaritons (DSPs) theory and thereby the developments of quantum
memory technique, i.e., transferring the quantum states of photon
wave-packet to collective Raman excitations in a loss-free and
reversible manner. All these works are based on a field theory
reformulation of the adiabatic approximation \cite{6}. In a very
recent paper, by extending the quantum state transfer technique to
propagating matter-waves, Fleischhauer and Gong formulated a
wonderful scheme to generate continuous beams of three-level
$\Lambda$ type atoms in non-classical or entangled states
\cite{7}. This new method may provide an experimentally accessible
way for the preparation of a continuous coherent atomic beam or an
atom laser that has drawn much attention in current literatures,
since the first pulsed atom laser was created in 1997 based on the
remarkable realization of Bose-Einstein condensation in dilute
atomic clouds \cite{8,9}. On the other hand, the the controlled
light storing in the medium composed of double $\Lambda$ type
four-level atoms was also investigated by several authors quite
recently \cite{10,11}. As a natural extension, in this paper, we
proceed to study the possible extending of the elegant
Fleischhauer-Gong scheme \cite{7} in the double $\Lambda$ atomic
configuration, in particular the transfer technique of quantum
states from a pair of probe pulses to matter waves. Besides, the
intensity of the output matter waves is found to be determined by
the initial relative phase of the two independent coherent probe
lights, which may indicate an interesting way to measure the
initial relative phase of two independent light sources.

We consider the quasi 1-dimensional system shown in Fig.1. A beam
of double $\Lambda$ type atoms interacts with two coherent probe
and two control Stokes fields and the former fields are taken to
be much weaker than the later. Atoms in different internal states
are described by four bosonic fields $\Psi_{\mu}(z,t)
(\mu=1,2,3,4)$. The Stokes fields can be described by the
Rabi-frequencies
$\Omega_j(z,t)=\Omega_{0j}(z)e^{-i\omega_{s_j}(t-z/c_j)}$ with
$\Omega_{0j}~(j=1,2)$ being taken as real and $c_j$ denoting the
phase velocities projected onto the $z$ axis, and the two coherent
probe fields are characterized by the dimensionless positive
frequency components
$E^{(+)}_j(z,t)=\varepsilon_j(z,t)e^{-i\omega_{pj}(t-z/c)}$.
Assuming $\omega_{p_1}-\omega_{s_1}=\omega_{p_2}-\omega_{s_2}$, we
may introduce the slowly-varying amplitudes, and a decomposition
into velocity classes
$\Psi_1=\sum_l\Phi^l_1e^{i(k_lz-\omega_lt)}$,
$\Psi_2=\sum_l\Phi^l_2e^{i[(k_l+k_{p_1})z-(\omega_l+\omega_{p_1})t]}$,
$\Psi_3=\sum_l\Phi^l_3e^{i[(k_l+k_{p_j}-k_{s_j})z-(\omega_l+\omega_{p_j}
-\omega_{s_j})t]}$, and
$\Psi_4=\sum_l\Phi^l_4e^{i[(k_l+k_{p_2})z-(\omega_l+\omega_{p_2})t]}$,
where $\hbar\omega_l=\hbar^2k^2_l/2m$ is the corresponding kinetic
energy in the $l$th velocity class, $k_{p_j} $ and $k_{s_j}$
$(j=1,2)$ are respectively the vector projections of the two probe
and Stokes fields to the $z$ axis. The atoms have a narrow
velocity distribution around $v_0=\hbar{k_0}/2m$ with
$k_0\gg|k_{p_j}-k_{s_j}|$, and all fields are assumed to be in
resonance for the central velocity class. Hence the equations of
motion for the atomic fields are given by
\begin{eqnarray}\label{eqn:1}
(\frac{\partial}{\partial{t}}+\frac{\hbar{k_l}}{2m}\frac{\partial}
{\partial{z}})\Phi^l_1
=-ig_1\varepsilon^*_1\Phi^l_2-ig_2\varepsilon^*_2\Phi^l_4
\end{eqnarray}
\begin{eqnarray}\label{eqn:2}
(\frac{\partial}{\partial{t}}+\frac{\hbar{k_l}}{2m}\frac{\partial}
{\partial{z}})\Phi^l_3
=-i\Omega_{01}\Phi^l_2-i\Omega_{02}\Phi^l_4-i\delta_l\Phi^l_3
\end{eqnarray}
\begin{eqnarray}\label{eqn:3}
(\frac{\partial}{\partial{t}}+\frac{\hbar(k_l+k_{p_1})}{2m}
\frac{\partial}{\partial{z}})\Phi^l_2
=-(\gamma_2+i\Delta_l)\Phi^l_2-i\Omega_{01}\Phi^l_3-ig_1\varepsilon_1\Phi^l_1+F^l_2
\end{eqnarray}
\begin{eqnarray}\label{eqn:4}
(\frac{\partial}{\partial{t}}+\frac{\hbar(k_l+k_{p_2})}{2m}
\frac{\partial}{\partial{z}})\Phi^l_4
=-(\gamma_4+i\Delta_l)\Phi^l_4-i\Omega_{02}\Phi^l_3-ig_2\varepsilon_2\Phi^l_1+F^l_4
\end{eqnarray}
where $g_j$ is the atom-field coupling constant between the states
$\lvec{1}$ and $\lvec{2}$ (for $j=1$) or $\lvec{1}$ and $\lvec{4}$
(for $j=2$), $\gamma_{2,4}$ denotes the loss rate out of the
excited state $\lvec{2}$ or $\lvec{4}$,
$\Delta_l\approx\hbar{k_l}k_{p1}/m+(\omega_{21}-\omega_{p1})\approx\hbar{k_l}k_{p2}/m+(\omega_{41}-\omega_{p2})$
and
$\delta_l\approx\hbar{k_l}(k_{pj}-k_{sj})/m+(\omega_{31}-\omega_{pj}-\omega_{sj})$
are the single and two-photon detunings and $F^l_{2,4}$ is the
corresponding Langevin noise operator that will be omitted in the
following derivation for simplicity. The propagation equations of
the two probe fields read
\begin{eqnarray}\label{eqn:5}
(\frac{\partial}{\partial{t}}+c\frac{\partial}{\partial{z}})\varepsilon_1(z,t)
=-ig_1\sum_l{{\Phi^{\dag}_1}^l\Phi^l_2}
\end{eqnarray}
\begin{eqnarray}\label{eqn:6}
(\frac{\partial}{\partial{t}}+c\frac{\partial}{\partial{z}})\varepsilon_2(z,t)
=-ig_2\sum_l{{\Phi^{\dag}_1}^l\Phi^l_4}
\end{eqnarray}

Consider a stationary input of atoms in state $\lvec{1}$, i.e.
$\Psi_1(0,t)=\sqrt{n}$, where $n$ is the constant total density of
atoms. In the limit of the two weak probe quantum fields and weak
atomic excitation one finds:\cite{7}
$\Phi^l_1(z,t)\approx\Phi^l_1(0,t-2mz/\hbar{k_l})=\sqrt{n}\xi_le^{-i\varphi_l(z,t)}$,
where $\sum_l{\xi_l}=1$ and $\varphi_l\equiv(k_lz-\omega_lt)$.
From the formula (\ref{eqn:3}) and (\ref{eqn:4}) one can find the
condition of the adiabatic evolution is fulfilled only in the
special case:
$\varepsilon_2(z,t)=\tan\vartheta(z)\varepsilon_1(z,t)$, where
$\vartheta$ is defined according to
$\tan\vartheta(z)=\frac{g_1\Omega_{02}(z)}{g_2\Omega_{01}(z)}$. In
general, however, the two input probe fields do not satisfy the
former special relation at the entrance region of the atoms. The
process is then nonadiabatic from the very beginning.
Raczy\'{n}ski et al.\cite{10} show that, due to the self-adjusting
of the two probe fields, the adiabatic condition will become
fulfilled after the nonadiabatic process, i.e.,
$\varepsilon_j(0,t-\tau(\delta))\rightarrow\varepsilon'_j(\delta,t)
$, where $\varepsilon'_2(z,t)=\tan\vartheta\varepsilon'_1(z,t)$
and $\tau(\delta)$ is the time shift from the entrance region to
an adjacent position $\delta$, $0<\delta<L$ with $L$ being the
interaction length in $z$ direction. To obtain the motion equation
of probe fields, in the region $\delta<z<L$, we definite a new
probe field: $
\varepsilon_{12}(z,t)=\cos\vartheta(z)\varepsilon'_1(z,t)+\sin\vartheta(z)\varepsilon'_2(z,t)
$, therefore we can obtain
\begin{eqnarray}\label{eqn:9}
(\frac{\partial}{\partial{t}}+c\frac{\partial}{\partial{z}})
\varepsilon_{12}(z,t)=-i\frac{g_1g_2}{\Omega}\sum_l{(\Omega_{01}\Phi^l_2(z,t)
+\Omega_{02}\Phi^l_4(z,t))\Phi^{\dag}_1}^l(z,t)
\end{eqnarray}
\begin{eqnarray}\label{eqn:10}
\Phi^l_3(z,t)&=&-\frac{g_1g_2}{\Omega}\varepsilon_{12}(z,t)\sqrt{n}
\xi_le^{-i\varphi_l(z,t)}+\frac{i}{\Omega_{01}+\Omega_{02}}[(\frac{\partial}{\partial{t}}
+\frac{\hbar(k_l+k_{p_1})}{2m}\frac{\partial}{\partial{z}}+\gamma_2+i\Delta_l)\Phi^l_2(z,t)\nonumber\\
&&+(\frac{\partial}{\partial{t}}+\frac{\hbar(k_l+k_{p_2})}{2m}
\frac{\partial}{\partial{z}}+\gamma_4+i\Delta_l)\Phi^l_4(z,t)]
\end{eqnarray}
where $\Omega=\sqrt{g^2_1\Omega^2_{02}+g^2_2\Omega^2_{01}}$. Here
we only consider the perfect two photon resonance case for all
atoms by omitting all the loss rates out of two excited states
(the validity of this approximation will be discussed later),
hence $\gamma_2=\gamma_4\equiv0$ and $\delta_l\equiv0$. In the
adiabatic approximation, we obtain
$\Phi^l_3(z,t)=-\frac{g_1g_2}{\Omega}\varepsilon_{12}(z,t)
\sqrt{n}\xi_le^{-i\varphi_l(z,t)}$ and
$\Omega_1\Phi^l_2(z,t)+\Omega_2\Phi^l_4(z,t)=-ig_1g_2\sqrt{n}\xi_le^{-i\varphi_l(z,t)}
(\frac{\partial}{\partial{t}}+\frac{\hbar{k_l}}{2m}
\frac{\partial}{\partial{z}})\frac{\varepsilon_{12}(z,t)}{\Omega(z)}$.
Substituting the latter results into the equation of the motion
for new probe field $\varepsilon_{12}(z,t)$ yields
\begin{eqnarray}\label{eqn:13}
[(1+\frac{g^2_1g^2_2n}{\Omega^2(z)})\frac{\partial}{\partial{t}}
+c(1+\frac{g^2_1g^2_2n}{\Omega^2(z)}\frac{v_0}{c})
\frac{\partial}{\partial{z}}]\varepsilon_{12}(z,t)
=\frac{g^2_1g^2_2n}{\Omega^2(z)}v_0(\frac{\partial}{\partial{z}}
\ln\Omega(z))\varepsilon_{12}(z,t)
\end{eqnarray}
with $v_0\equiv\sum_l{\xi_l}v_l$. The r.h.s. of this equation
describes a reduction (enhancement) due to stimulated Raman
adiabatic passage in two spatially varying Stokes fields for
$v_0\neq0$, and the space-dependent Stokes fields in laboratory
frame is equivalent to the time-dependent fields in the rest frame
of the atoms. Introducing the mixing angle $\theta(z)$ according
to
$\tan^2\theta(z)\equiv\frac{g^2_1g^2_2n}{\Omega^2}\frac{v_0}{c}$,
one can easily find the solution:
$\varepsilon_{12}(z,t)=\varepsilon_{12}(\delta,t-\tau(z,\delta))
\frac{\cos\theta(z)}{\cos\theta(\delta)}$, where
$\tau(z,\delta)=\tau(z)-\tau(\delta)=\int^z_{\delta}{dz'V^{-1}_g(z')}$
with the group velocity
$V_g=c(1+\frac{g^2_1g^2_2n}{\Omega^2}\frac{v_0}{c})/(1
+\frac{g^2_1g^2_2n}{\Omega^2})$ that approaches $v_0$ for
$\Omega(z)\rightarrow0$. The initial value of $\varepsilon_{12}$
can be calculated as\cite{10}:
$\varepsilon_{12}(\delta,t)=(\cos\vartheta(0)\varepsilon_1(0,t-\tau(\delta))
+\sin\vartheta(0)\varepsilon_2(0,t-\tau(\delta)))\cos\theta(\delta)/\cos\theta(0)$.
Assuming  $\theta(0)=0$ and  $\theta(L)=\pi/2$ at the input and
output points respectively, the bosonic field $\Phi_3(z,t)$ is
then obtained by
\begin{eqnarray}\label{eqn:14}
\Phi_3(z,t)=-\sqrt{\frac{c}{v_0}}(\cos\vartheta_0\varepsilon_1(0,t-\tau(L))
+\sin\vartheta_0\varepsilon_2(0,t-\tau(L)))
\end{eqnarray}
where $\tau(L)=\tau(\delta)+\int^L_{\delta}{dz'V^{-1}_g(z')}$ and
$\vartheta_0=\vartheta(0)$. The factor $\sqrt{c/v_0}$ accounts for
the fact that the input light pulses propagate with velocity $c$
and the output matter field propagates with $v_0$. Furthermore
\begin{eqnarray}\label{eqn:15}
v_0|\Psi_3|^2_{out}=c(\cos^2\vartheta_0|\varepsilon_1|^2_{in}
+\sin^2\vartheta_0|\varepsilon_2|^2_{in}+2Re({\cos\vartheta_0\sin\vartheta_0\varepsilon^{*}_1\varepsilon_2})_{in})
\end{eqnarray}
Noting the complex amplitudes
$\varepsilon_j=|\varepsilon_j|e^{i\phi_j}~(j=1,2)$, where $\phi_j$
corresponds to the initial phase of the probe lights, the last
term of the r.h.s of above formula then recasts into:
$2\cos\vartheta_0\sin\vartheta_0|\varepsilon_1|_{in}|\varepsilon_2|_{in}\cos\phi$,
where $\phi=\phi_1-\phi_2$ is the initial relative phase, and it
gives a crucial influence on the intensity of output
$\Phi_3$-atomic beam. As a result of the non-adiabatic evolution,
from Eq.(\ref{eqn:15}) one can easily find the output flux of
$\Phi_3$-atoms is less than (or equal) the total input flux of
photons, i.e. $v_0|\Psi_3|^2_{out}\leq
c(|\varepsilon_1|^2_{in}+|\varepsilon_2|^2_{in})$. Particularly,
(a) if $g_2\Omega_{01}(0)\gg{g_1}\Omega_{02}(0)$, we have
$\cos\vartheta(0)\simeq1$ and
$v_0|\Psi_3|^2_{out}=c|\varepsilon_1|^2_{in}$, which indicates the
output flux of $\Phi_3$-atoms equals to the input flux of the
photons of the probe field $\varepsilon_1$, while all photons of
$\varepsilon_2$ is damped in the non-adiabatic process from the
very beginning; (b) If $g_2\Omega_{01}(0)\ll{g_1}\Omega_{02}(0)$,
we have $\cos\vartheta(0)\simeq0$ and
$v_0|\Psi_3|^2_{out}=c|\varepsilon_2|^2_{in}$, which indicates the
output flux of $\Phi_3$-atoms equals to the input flux of the
photons of the probe field $\varepsilon_2$,while $\varepsilon_1$
is damped in the non-adiabatic process. These results indicate
that the quantum states of the continuous output matter waves can
easily be steered by the two control stokes fields.

The formula (\ref{eqn:15}) is the main result in this paper. For
this we may present a possible application for the measurement of
the initial relative phase $\phi$ of two independent coherent
lights. The schematic measurement setup is shown in Fig.2. M is a
semitransparent mirror splitter, through which the two input probe
pulses $E^0_1(z,t)$ and $E^0_2(z',t)$ are split into four pulses
with identical intensities, i.e. $E_1(z,t)$, $E_2(z,t)$,
$E'_1(z't)$ and $E'_2(z',t)$ with their amplitudes
$|\varepsilon_1|=|\varepsilon_2|=|\varepsilon'_1|=|\varepsilon'_2|=\varepsilon_0$.
Among these splitters, $E'_1(z',t)$ and $E'_2(z',t)$ enter the (-)
channel, while $E_1(z,t)$ and $E_2(z,t)$ enter the (+) one. G is a
plate glass with its thickness $d=\pi/|(n-1)(k_{p1}-k_{p2})|$,
where $n$ is the refractive index of the two probe pulses in the
glass. The relative phase between $E_1(z,t)$ and $E_2(z,t)$
increases an additional value $\pi$ after passing through the
glass G. Assuming that $g_1\Omega_2(0)={g_2}\Omega_1(0)$, the
intensities of the outputs of $\Phi_3$-atom flux from channel (+)
and (-) are then given by
\begin{eqnarray}\label{eqn:17}
I^+_3=I_0\sin^2\frac{\phi}{2},&& I^-_3=I_0\cos^2\frac{\phi}{2}
\end{eqnarray}
where $I_3=|\Psi_3|^2_{out}$ and
$I_0=\frac{v_0}{c}\varepsilon_0^2$. The recording of $I_3^{\pm}$
allows one to determine the absolute value of the relative phase
$\phi$ of two independent coherent probe fields. Note that $\phi$
is an unpredictable random variable, which takes a different value
for any new realization. We consider the case of $k_{\pm}$
detected $\Phi_3$-atoms in the ($\pm$) channel for a fixed number
of measurements $k=k_{+}+k_-$. Each count occurs with
probabilities $\sin^2\frac{\phi}{2}$ and $\cos^2\frac{\phi}{2}$ in
the (+) and (-) channels. Note $P(k_+,k_-,\phi)$ as the
probability for the result ($k_+,k_-$), when $k\gg1$, one can
easily find $P(k_+,k_-,\phi)$ is maximal for
$\phi=2\tan^{-1}\sqrt{k_+/k_-}$, and the shot noise on the signal
in the two channels ($\pm$) can be neglected \cite{12}. The phase
$\phi$ can be then well determined by the ratio $k_+/k_-$ detected
on the counters.

Now we discuss in detail on the approximations invoked in the
derivation of the above results. First, from the relation between
$\Phi_3$ and $\varepsilon_{12}(z,t)$ one finds
$g^2_1g^2_2\langle\varepsilon^{\dag}_{12}\varepsilon_{12}\rangle/\Omega^2
=\langle\Psi^{\dag}_3\Psi_3\rangle/n$. Then, as long as the
condition of weak atomic excitation fulfills, in other words, when
the input flux of atoms is much larger than the input flux of pump
photons, the adiabatic approximation used in above discussion is
valid. Another approximation is that we have assume the perfect
two photon resonance. For the case of a non-vanishing  but
constant value of $\delta_l=\delta$ and equal loss rates
$\gamma_2=\gamma_4=\gamma$, the accompanying contribution to
$\Omega_{01}\Phi^l_2+\Omega_{02}\Phi^l_4$ in the lowest order is
\begin{eqnarray}\label{eqn:19}
\Omega_{01}\Phi^l_2+\Omega_{02}\Phi^l_4\rightarrow\Omega_{01}\Phi^l_2
+\Omega_{02}\Phi^l_4+\frac{(\Omega^2_{01}+\Omega^2_{02})\delta}{\Omega^2_{01}
+\Omega^2_{02}-\delta\Delta+i\gamma\delta}\frac{g_1g_2\sqrt{n}}{\Omega}
\varepsilon_{12}\xi_le^{-i(k_lz-\omega_lt)}
\end{eqnarray}
Substituting above result to the formula (\ref{eqn:9}), one can
easily see that the additional imaginary (real) terms of the
formula(\ref{eqn:19}) bring a loss (a phase shift) of
$\varepsilon_{12}$, i.e.
$\varepsilon_{12}\rightarrow{e^{-\alpha_1+i\alpha_2}\varepsilon_{12}}$
with $\alpha_i(i=1,2)$ real, and then the output $\Phi_3$-atom
flux intensity: $I^{\pm}_3\rightarrow{e^{-2\alpha_1}I^{\pm}_3}$.
For $\Delta=0$, the parameter $\alpha_{1,2}$ can be calculated by
$\alpha_1=\int^L_0{dz\sin^2\theta(z)\delta^2\gamma(\Omega^2_{01}+\Omega^2_{02})/(v_0((\Omega^2_{01}+\Omega^2_{02})^4+\delta^2\gamma^2))}$
and
$\alpha_2=\int^L_0{dz\sin^2\theta(z)(\Omega^2_{01}+\Omega^2_{02})^2\delta^2/(v_0((\Omega^2_{01}+\Omega^2_{02})^4+\delta^2\gamma^2))}$.
Since the detected value of $k_+/k_-$ depends on the ratio between
$I^+_3$ and $I^-_3$ instead of on $I^+_3$ or $I^-_3$ only, the
above derivation for the measurement of relative phase is also
valid. On the other hand, to obtain a high efficiency in the
transfer technique of quantum state from probe fields to matter
waves, the value of $\alpha_1$ should be small enough. Generally,
we may assume that $g_1\geq{g_2}$, one can then obtain
$\alpha_1\leq{\eta\int^1_0{d\xi\frac{\cos^2\beta(\xi)\sigma^2}{\cot^4\beta(\xi)+\sigma^2}}}$,
where $\beta$ is defined via
$tan^2\beta=\frac{g^2_1n}{\Omega^2_{01}+\Omega^2_{02}}\frac{v_0}{c}$,
$\xi=z/L$, $\sigma\equiv\delta\gamma/g^2_1n\frac{v_0}{c}$ and
$\eta\equiv{g^2_1nL/\gamma{c}}$ is the opacity of the medium in
the absence of EIT. The following discussions is then similar as
that in ref.\cite{7}. Assuming $\sigma\ll1$ one can give an upper
limit to the integral as
$\int^1_0{d\xi\frac{\cos^2\beta(\xi)\sigma^2}{\cot^4\beta(\xi)+\sigma^2}}\leq|\sigma|/2$,
consequently $\alpha_1\leq\eta|\sigma|/2$. Thus the case
$|\delta|L/v_0\ll1$ meets the request of a small value of
$\alpha_1$. A residual Doppler shift of the $1-3$ transition can
result in a two-photon detuning through $\delta_j=\Delta
v(\vec{k}_{pj}-\vec{k}_{sj})\cdot\vec{e}_z(j=1,2)$, where
$\Delta{v}$ denotes the difference of the velocity in $z$
direction with respect to the resonant velocity class. Then the
condition for a small $\alpha_1$ can also reads $\Delta{v}/v_0\ll
min \{1/(k_{pj}-k_{sj})L \}$.

In conclusion we have extended the transfer technique of quantum
states from a pair of probe fields to collective atomic excitation
to matter-waves in a double $\Lambda$ type system. We observe that
the quantum states of the continuous output matter-wave beam can
be steered by the two control stokes fields, and their intensity
can be determined by the initial relative phase of the lights. For
this an interesting potential application in measuring the initial
relative phase of two independent coherent sources is also briefly
discussed. Taking into account of present difficulties in making a
continuous atom laser in laboratory based on either the output of
a trapped Bose condensate with finite atoms number or the moving
magneto-optic trap and evaporative cooling technique \cite{13},
the problem studies in this paper raises an interest since it
confirms the elegant Fleischhauer-Gong scheme \cite{7} in this
more general configuration also as a novel way in next generation
of atom-laser experiments.

We thank Xu Jingjun, Zheng-Xin Liu, and Guimei Jiang for valuable
discussions. This work is supported by NSF of China under grants
No.10275036 and No.10304020. One of the authors (H. J.) is also
supported by CAS K.-C. Wong education Fund (HongKong), China
Postdoctoral Science Foundation and Postdoctoral Research Plan in
Shanghai.



\noindent\\  \\

\begin{figure}[ht]
\includegraphics[width=0.65\columnwidth]{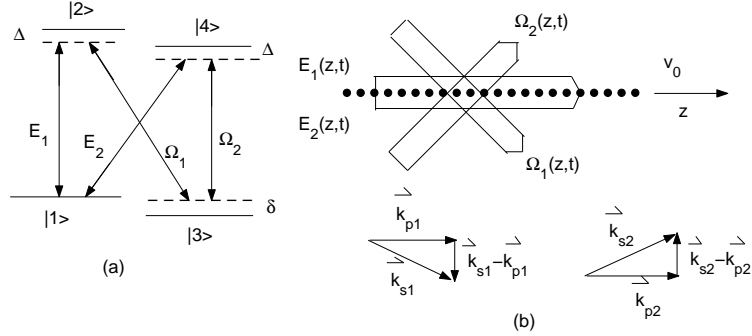}
\caption{(a)Beam of double $\Lambda$ type atoms coupled to two
control fields and two coherent probe fields. (b)To minimize
effect of Doppler-broadening, geometry is chosen such that
$(\vec{k}_{pi}-\vec{k}_{si})\cdot\vec{e}_z\approx0$ $(i=1,2)$.}
\label{}
\end{figure}

\begin{figure}[ht]
\includegraphics[width=0.6\columnwidth]{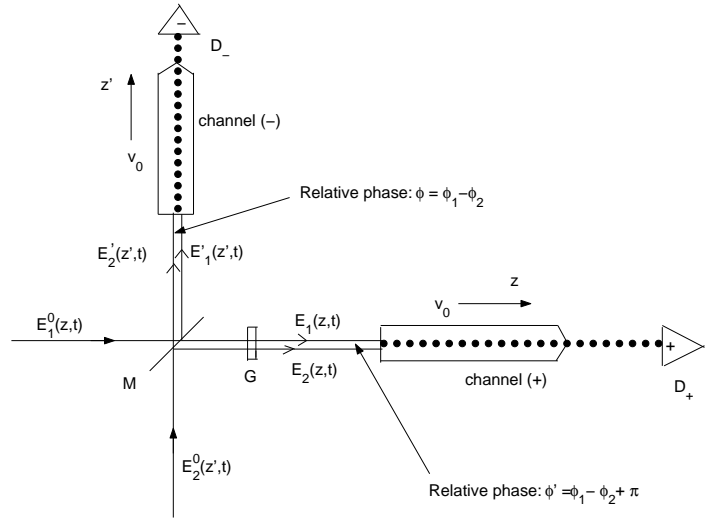}
\caption{The setup schematic for measurement. Here the geometry
relation between the probe and control fields is the same as that
in fig.1(b), for simplicity which is not shown in this figure. The
output $\Phi_3$-atoms are detected on $D_{+}$ and $D_{-}$, from
which the relative phase between $\varepsilon_1$ and
$\varepsilon_2$ can be measured. }\label{}

\end{figure}
\end{document}